\documentstyle[12pt]{article}
\begin{document}
\def\Ds{D\mkern-11.5mu/\,}  
\title{Critical dimension of Spectral Triples}

\author{Alejandro RIVERO \thanks{EUPT, 
Univ de Zaragoza, Campus de Teruel, 
 44003 Teruel, Spain} \thanks{email: arivero@unizar.es}}

\maketitle

\begin{abstract}
It is open the possibility of imposing requisites to the quantisation
of Spectral Triples in such a way that a critical dimension
D=26 appears.
\end{abstract}

From \cite{connes96} it is known that commutative spectral
triples contain the Einstein Hilbert action, which is
extracted by using the Wodziski residue over $\Ds^{-2} |\Ds|^n$,
being $\Ds$ a Dirac operator.

The theorem was initially enunciated  \cite{connes96} with
a complicated proportionality factor,
$$
c_n= {n-2 \over 12} {1 \over (4\pi)^{n/2}}
     {1\over \Gamma(n/2 +1)} 2^{[n/2]}
$$
and initial proofs where given independently by Kastler, and by
Kalau and Walze. There it was already noticed that the normalisation 
of the residue was somehow arbitrary. Actually, the factor in
the previous expression contains three elements:

-a volume $\Omega_n$ of the n-dimensional sphere

-the dimension $2^{[n/2]}$ of the fiber of the Dirac operator.

-an extant term $n-2 \over 24$ {\big (!!!)}

Further development of the theory has driven to include the two
first elements in the normalisation of the residue, so that
the definition coincides with the generic integral over a non
commutative manifold.
 Besides, modern proofs of the theorem get the extra factor
in a very independent way: during a expansion of the
spectral density kernel, 
it comes \cite[eq. 8.12]{varilly}from a coefficient 
$ {n-2 \over 2 }(\frac16 r(x)-c(x))$
when one applies Lichnerowicz formula, $c=\frac14 r$. 

Thus the right normalised form of the fundamental theorem
for commutative spectral triples has evolved to show explicitly
the extra factor. For instance already \cite[pg. 33]{martinetti} enounces
$$
S(\Ds)= - {n-2 \over 24} \int_M s  \sqrt{\det g} \; d^nx
$$
(the minus sign still there 
is due to Euclidean Gravity, which is the 
usual formulation in the context of non commutative geometry)
The change in normalisation can be traced back to 
 \cite[Th. 11.2]{book}, in a book which is to NCG theory as 
Polchinski's is to string theory.

Coming naturally from the mathematical development, it seems that nobody
has raised any issue about this extant factor.
But it is evident that if we 
require it to be equal to one,
we are imposing space time dimension n=26, an amusing coincidence.
Now, is there some situation where the extant factor can be
required to be unity? 
Surely some different ways can be
found to impose this requeriment, as it happened during
the evolution of string theory. 
 
Our first suggestion is that
a perturbative quantisation of the action given by
the normalised trace of the spectral triple will accumulate powers
of this term, so it will coincide with perturbative 
quantisation of gravity only when the factor is 1.  
Thus we claim that a perturbative quantisation of
commutative spectral triples gives gravity only if the
dimension of the triple is n=26.

At the moment we have not argument to tell that
quantisation in other number of dimensions is inconsistent.
This is slightly different from the proofs in string 
theory, where consistency implies gravity and at the same
time consistency requires 26 dimensions.

Nature has given us, up to today, sort of 24 degrees of freedom: 12 
elementary fermions and 12 gauge bosons-. Bosonic strings have, 
of course, 24 transverse directions, but no fermions. Heterotic 
strings have a narrow miss
trying to score this target, but model builders opt towards a not 
straightforward 
implementation of the Standard Model spectrum.  Having another
theory with criticality at D=26 could guide us towards the
M[issing] theory. Let us point out that in NCG the Higgses are
very coordinate-like, so perhaps they can be counted
jointly with space-time coordinates in order to partner the
fermions.

The intriguing point in taking a NCG approach is that we are not
using any tool from string theory. Thus
we confront two alternatives: either the existence of
a critical dimension D=26 is a phenomena independent of
string theory, or commutative spectral triples can be
obtained as a limit of bosonic strings. This second possibility
is very interesting because it reopens the door of directly
studying [bosonic?] strings in the context of noncommutative
spectral triples. Formulation of spectral triples associated
to string interactions had already been done in the past,
before the flood of papers on non commutativity. Check 
\cite{lizzi} for an instance. 

In is tempting to speculate if and how could the almost commutative
limit of a concrete string theory give not only the justification
of the critical factor, but also the 0-dim part of an almost
commutative spectral triple, ie the Connes-Lott model.

\end{document}